\documentclass[]{interact}
\pdfoutput=1

\usepackage{epstopdf}
\usepackage[caption=false]{subfig}

\usepackage[numbers,sort&compress]{natbib}
\bibpunct[, ]{[}{]}{,}{n}{,}{,}
\renewcommand\bibfont{\fontsize{10}{12}\selectfont}

\theoremstyle{plain}

\theoremstyle{definition}

\usepackage{url}
\usepackage{mathtools}
\usepackage{hyperref}

\newcommand\diff{\mathrm{d}}
\usepackage{braket}
\newcommand{\imag}{\mathrm{i}}

\usepackage[usenames,dvipsnames]{xcolor}

\begin{document}


\title{Mode-coupling theory of the glass transition for colloidal liquids in slit geometry}
\author{
\name{Lukas Schrack\textsuperscript{a} and Thomas Franosch\textsuperscript{a}}
\affil{\textsuperscript{a}Institut f\"ur Theoretische Physik, Universit\"at Innsbruck, Technikerstr. 21A, 6020 Innsbruck, Austria}
}
\maketitle

\begin{abstract}
We provide a detailed derivation of the mode-coupling equations for a colloidal liquid confined by two parallel smooth walls. We introduce irreducible memory kernels for the different relaxation channels thereby extending the projection operator technique to colloidal liquids in slit geometry. Investigating both the collective dynamics as well as the tagged-particle motion, we prove that the mode-coupling functional assumes the same form as in the Newtonian case corroborating the universality of the glass-transition singularity with respect to the microscopic dynamics. 
\end{abstract}


\section{Introduction}
Confined liquids are intermediate between a three-dimensional (3D) bulk  and a quasi-two-dimensional (2D) system, thereby introducing a competition between the near-range local ordering and the constraints imposed by the boundary. The confinement, in the simplest case a slit geometry of two parallel smooth walls, introduces a new control parameter into the problem giving rise to plethora of new structural and dynamic phenomena~\cite{Loewen:JPCM:2001,Alba:JoP:2006,Varnik:JP:2016}, including the layering of the average density~\cite{Nugent:PRL:2007}, complex intermediate phases due to stacking~\cite{Schmidt:PRL:1996, Schmidt:PRE:1997}, and a shift in the glass transition~\cite{Nugent:PRL:2007,Eral:PRE:2009,Edmond:Eur:2010,Ingebrigsten_PRL:2013}. The slit geometry then permits to study the dimensional crossover as the plate separation becomes small and only recently the decoupling of the in-plane degrees of freedom from the transverse ones has been unravelled to make systematic expansion for the structural properties~\cite{Franosch:PRL:2012, Lang:JCP:2014} as well as for the dynamics~\cite{Schilling:PRE:2016,Mandal:PRL:2017, Mandal:EPJST:2017}.

A striking empirical observation is that diffusion correlates with purely thermodynamic quantities such as the excess entropy and the average fluid density~\cite{Rosenfeld:PRA:1977,Dzugutov:Nature:1996,Rosenfeld:JPCM:1999}.
This correlation even extends to confined liquids~\cite{Mittal:PRL:2006,Mittal:PRL:2008,Goel:PRL:2008,Goel:JStatMech:2009,Bollinger:JCP:2015} which  suggests that microscopic details enter transport properties only indirectly via structural quantities, corroborating an engineer's or biologist's paradigm: structure determines function, where function in this context means dynamic behaviour. 

In a related spirit the microscopic approach as formulated in the mode-coupling theory of the glass transition (MCT)~\cite{Goetze:Complex_Dynamics} derives a closed set of equations of motion for 
time-dependent correlation functions such that only measurable structural information is required as input. The theory makes a series of non-trivial predictions for the universality of the glass transition which have been tested successfully in numerous experiments and computer simulations for bulk systems~\cite{Goetze:Complex_Dynamics, Janssen:FiP:2018}. For confined systems a symmetry-adapted MCT has been elaborated more recently~\cite{Lang:PRL:2010, Lang:PRE_90:2014, Lang:PRE_89:2014, Lang:PRE:2012} for Newtonian dynamics. In particular,  this theory predicts a reentrant scenario for the glass-transition line which has been corroborated also in event-driven molecular dynamics simulations~\cite{Mandal:NatComm:2014, Mandal:SoftMatter:2017}. An interesting variant of MCT for confinement on  a hypersphere to elucidate the role of curvature and frustration has also been investigated recently~\cite{Vest:JCP_143:2015, Vest:JCP_138:2018, Turci:PRL_118:2017}. 

Experimentally the slit geometry is difficult to realize for molecular liquids since the confining walls will display a roughness on the molecular scale. Yet, for colloidal systems the separation of length scales allows to use glass plates which appear smooth on the scale of the colloid diameters. The structural properties of such confined colloidal systems have been explored experimentally by X-ray scattering \cite{Nygard:PRL:2012,Nygard:JCP:2013}, clearly demonstrating the anisotropies and layering induced by the walls. Moreover, the changes in the short-time dynamics as manifested by the de Gennes narrowing have become experimentally accessible by scattering techniques~\cite{Nygard:PRL:2016}. The ramification of confinement on colloidal diffusion have only recently become a focus of experimental investigations, see Ref.~\cite{Nygard:PCCP:2017} for a review.   

Changing the microscopic dynamics from Newtonian to Brownian does not change the glassy dynamics for bulk systems as predicted by MCT~\cite{Szamel:PRA:1991,Goetze:Complex_Dynamics,Franosch:JNCS:1998,Mandal:SoftMatter:2018}. For the case of confinement the universality of the glassy dynamics has been anticipated implicitly by comparing the non-equilibrium state diagram of the MCT with Newtonian dynamics~\cite{Lang:PRL:2010} to  experiments on colloidal systems~\cite{Mittal:PRL:2008}. Furthermore, the equations of motion for the colloidal case have been conjectured~\cite{Lang:JStatMech:2013} in analogy to the bulk system, yet a microscopic derivation is still missing. 

The goal of this work is to provide a step-by-step derivation  of the MCT equations in confinement for  the colloidal case. We shall demonstrate that the Zwanzig-Mori projection operator scheme can be adapted such that the force correlation kernels of a Newtonian fluid are replaced by suitable irreducible memory kernels also for the natural splitting of relaxation modes. These formally exact representations of the new memory kernel then provides the starting point for the mode-coupling approximation. Using the standard techniques, we show that the MCT memory kernels assume the same form as in the Newtonian case corroborating the universality of glassy dynamics also in confinement.

\section{Dynamics in confinement}
For the dynamic properties of a confined colloidal liquid we first recall  the basic formalism of time-dependent  correlation functions in equilibrium driven by Brownian dynamics. Then we introduce the observables of fundamental interest and use the symmetries of the confinement geometry to obtain a proper mode decomposition. As a result, all quantities of interest can be expressed in terms of suitably generalized matrix-valued intermediate scattering functions.  

\subsection{Model and general formalism}
We consider a simple colloidal liquid  comprised of $N$ identical spherical particles in suspension.  It is enclosed between two flat, hard and parallel walls of area $A$ such that the centres of the particles are confined to $|z|\le L/2$. The liquid is considered to be in thermal equilibrium at temperature $T$. The thermodynamic limit $N\to\infty, A\to \infty$ is anticipated throughout such that the area density $n_0=N/A$ remains constant. The set-up includes the case of hard exclusion interactions between the particles and the walls, then the physical separation of the plates is $H= L +\sigma$, where $\sigma$ is the diameter of the hard spheres.

The in-plane or lateral coordinates are denoted by $\vec{r}=(x,y)$, the perpendicular or transversal ones  by $z$, together we use the short-hand notation $\vec{x} = (\vec{r}, z)$ for a position in the slit. The collection of the positions of the centres of the particles $\Gamma = (\vec{x}_1,\ldots , \vec{x}_N)$ defines a point in $3N$-dimensional configuration space. The dynamics of the colloidal suspension will be treated as overdamped and hydrodynamic interaction will be ignored. Then the dynamics of the conditional probability density $\Psi(\Gamma, t| \Gamma', 0)$ to find the system in a configuration $\Gamma$ at time $t$ provided it started from configuration $\Gamma'$ at initial time $t^\prime=0$ is driven by the Smoluchowski equation~\cite{Dhont:Introduction_to_Dynamics_of_Colloids, Doi:Oxford:1999}
\begin{equation}\label{eq:Smoluchowski}
 \partial_t\Psi=\Omega\Psi,
\end{equation}
with the Smoluchowski operator
\begin{equation}
 \Omega = \Omega(\Gamma) =  D_0\sum_{n=1}^N\vec{\nabla}_n\cdot \left(\vec{\nabla}_n+\frac{1}{k_B T}\vec{\nabla}_n U\right),
\end{equation}
acting on the configuration $\Gamma$. Here $D_0$ denotes the bare diffusion coefficient, $k_B T$ is  the thermal energy, and the potential energy $U = U(\Gamma)$ includes the interactions among the colloids as well as the interaction with the walls. The equilibrium distribution is then provided by the canonical ensemble
\begin{equation}\label{eq:equilibrium_distribution}
 \psi_{\text{eq}}  (\Gamma) = Z^{-1} \exp[ - U(\Gamma )/k_B T],
\end{equation}
which fulfils $\Omega \psi_{\text{eq}} = 0$. 

General time-dependent correlation functions between observables $A=A(\Gamma), B= B(\Gamma)$ are defined by~\cite{Dhont:Introduction_to_Dynamics_of_Colloids} 
\begin{equation}
C_{AB}(t) := \int \diff \Gamma \diff \Gamma'  A(\Gamma)^* \Psi(\Gamma, t| \Gamma', 0) B(\Gamma') \psi_{\text{eq}}(\Gamma'),
\end{equation}
where the integrals are over the initial and final configuration space. This definition reflects directly the experimental protocol of correlating the observables at different instances of time, in particular it reduces to canonical  equal-time averages for $t=0$. The formal solution of the Smoluchowski equation, Eq.~\eqref{eq:Smoluchowski}, $\Psi(\Gamma, t| \Gamma', 0) = \exp( \Omega t) \delta(\Gamma,\Gamma')$ can be used to map the time dependence from the probabilities (Schr\"odinger picture) to the observables (Heisenberg picture) 
\begin{equation}
C_{AB}(t) = \int \diff \Gamma \left[ e^{\Omega^\dagger t} A(\Gamma) \right]^* B(\Gamma) \psi_{\text{eq}}(\Gamma),
\end{equation}
where the adjoint Smoluchowski operator reads
\begin{equation}\label{eq:Smoluchowski_adjoint}
\Omega^\dagger=D_0\sum_{n=1}^N (\vec{\nabla}_n-\frac{1}{k_B T}\vec{\nabla}_n U) \cdot \vec{\nabla}_n.
\end{equation}
Then it is natural to define time-dependent observables  
\begin{equation}
 A(t) \equiv A(\Gamma,t) := \mathcal{R}(t) A(\Gamma)  ,
\end{equation}
with the time-evolution operator $ \mathcal{R}(t)=\exp{\left(\Omega^\dagger t\right)}$,
as well as the Kubo scalar product~\cite{Goetze:Complex_Dynamics}
\begin{equation}\label{eq:Kubo_scalar}
 \langle A | B \rangle := \int \diff \Gamma\, \delta A(\Gamma)^* \delta B(\Gamma) \psi_{\text{eq}}(\Gamma),
\end{equation}
in the Hilbert space of (fluctuating) observables $\delta A := A - \langle A \rangle$. 
Accordingly the dynamic correlation function
\begin{align}
 C_{AB}(t) 
=\braket{A(t)|B}=\braket{A|\mathcal{R}(t)|B},
\end{align}
is merely  the matrix element of the time-evolution operator with respect to the bra- $\langle A |$ and ket-state $| B \rangle$. The bra-ket notation for the Hilbert space structure will be the starting point for the projection operator technique by Zwanzig and Mori~\cite{Hansen:Theory_of_Simple_Liquids,Goetze:Complex_Dynamics} to derive formally exact equations of motion for the dynamic correlation functions in terms of memory kernels. 

For a set of observables $A_i$, the time-dependent matrix $\langle A_i(t) | A_j \rangle$  fulfils the general properties of matrix-valued autocorrelation functions~\cite{Feller:Probability,Gesztesy:Math:2000,Lang:JStatMech:2013}. In particular, for any set of complex numbers $y_i$ the contraction $\sum_{ij} y_i^* \langle A_i(t) | A_j \rangle y_j$ is the  autocorrelation function of the variable $A(t) = \sum_i y_i A_i(t)$. By the spectral representation theorem~\cite{Feller:Probability} it corresponds to the characteristic function $\int e^{-\imag \omega t} R(\diff \omega)$  of a  finite (symmetric) Lebesgue-Stieltjes measure $R(\omega)$. 
 
For the case of Brownian motion even stronger statements can be made. One readily checks that the adjoint Smoluchowski operator is hermitian with respect to the Kubo scalar product $\langle A | \Omega^\dagger B \rangle = \langle \Omega^\dagger A | B \rangle$ and that $\langle A | \Omega^\dagger A \rangle \leq 0$.  Therefore the spectrum of $-\Omega^\dagger$ is real and non-negative.  
Then the time-evolution operator admits a representation as ${\cal R}(t) = \int_0^\infty e^{-\gamma t} \diff E(\gamma) $ with a projection-valued measure $E(\diff \gamma)$. This implies that  autocorrelation functions are expressed as Laplace transforms 
\begin{align}\label{eq:Bernstein}
C_{AA}(t) =  \langle A(t)| A \rangle = \int_0^\infty e^{-\gamma t} a(\diff\gamma),
\end{align}
of a  finite Lebesgue-Stieltjes measure $a(\diff \gamma) := \langle A | E(\diff \gamma) A \rangle$. In particular, this implies that autocorrelation functions are \emph{completely monotone}, i.e. they change sign upon taking subsequent  time-derivatives $ [-\partial_t]^\ell C_{AA}(t) \geq 0, \ell \in \mathbb{N}$. By Bernstein's theorem~\cite{Feller:Probability} this condition is equivalent to the representation in Eq.~\eqref{eq:Bernstein}. Especially, completely monotone functions are a subclass of the class of correlation functions. Generalizations to the matrix-valued case are straightforward by taking the contractions $\sum_{ij} y_i^* \langle A_i(t) | A_j \rangle y_j$ as above.

\subsection{Correlation functions in confinement} 
The following subsection follows mainly the presentation of Ref.~\cite{Lang:PRE:2012}  and serves to fix the notation adapted to the case of Brownian dynamics and to  make the paper self-contained. 

The simplest observable characterizing the fluid is the local fluctuating density, defined microscopically by 
\begin{equation}
 \rho(\vec{r},z)=\sum_{n=1}^N\delta[\vec{r}-\vec{r}_n]\delta[z-z_n].
\end{equation}
Due to translational symmetry parallel to the confining, the equilibrium density $n(z)$ only depends on the perpendicular coordinate $z$
\begin{equation}
 n(z) = \braket{\rho(\vec{r},z)},
\end{equation}
where $\braket{\dots}$ describes canonical averaging. We introduce fluctuations
\begin{equation}
 \delta\rho(\vec{r},z) \coloneqq \rho(\vec{r},z) - n(z),
\end{equation}
and define the time-dependent density-density correlation function corresponding to the Van Hove function~\cite{Hansen:Theory_of_Simple_Liquids,Henderson:Fundamentals_of_inhomogeneous_fluids}
\begin{equation}
 G(|\vec{r}-\vec{r}^{\,\prime}|,z,z^\prime,t) \coloneqq \frac{1}{n_0}\braket{ \delta
  \rho(\vec{r},z,t) \delta\rho(\vec{r}^{\,\prime},z^\prime) }.
\end{equation}
Due to translational symmetry parallel to the planes and rotational symmetry around an axis perpendicular to the plane, the Van Hove function
only depends on the magnitude of the relative lateral positions $|\vec{r}-\vec{r}^{\,\prime}|$ and explicitly on both transversal positions $z,z'$. For our case the walls are symmetric, such that a simultaneous sign change of the transversal coordinates, $z \mapsto - z, z' \mapsto -z'$, does not change the Van Hove function.

We expand the dependence on the transversal positions in all quantities in terms of discrete Fourier modes $\exp{\left(\imag Q_\mu z\right)}$ in $z-$direction ($Q_\mu=2\pi\mu/L, \mu\in\mathbb{Z}$). In contrast, the spatial dependence parallel to the surface is decomposed into ordinary plane waves $\exp(-\imag \vec{q}\cdot\vec{r})$ with $\vec{q}=(q_x,q_y)$. These wave vectors are taken as discrete initially $(q_x, q_y) \in (2\pi/\sqrt{A}) \mathbb{Z}^2$, however, in the thermodynamic limit they become continuous variables, such that sums are replaced by integrals $(1/A) \sum_{\vec{q}} \ldots \mapsto (2\pi)^{-2} \int \diff^2 q \ldots$ as usual. Then the following  orthogonality and completeness relations hold
\begin{align}
 &\frac{1}{A}\int_A \mathrm{d}\vec{r} e^{\imag(\vec{q}-\vec{q}^{\,\prime})\cdot\vec{r}} =
  \delta_{\vec{q},\vec{q}^{\,\prime}}, \\
 &\frac{1}{A}\sum_{\vec{q}}e^{\imag \vec{q}\cdot(\vec{r}-\vec{r}^{\,\prime})} = 
  \delta(\vec{r}-\vec{r}^{\,\prime}), \\
 &\frac{1}{L}\int_{-L/2}^{L/2}\mathrm{d}z \exp{\left[\imag(Q_\mu-Q_{\mu^\prime})z\right]} = 
  \delta_{\mu\mu^\prime}, \\
 &\frac{1}{L}\sum_\mu\exp{\left[\imag Q_\mu (z-z^\prime)\right]} = 
  \delta(z-z^\prime).
\end{align}
In particular, the equilibrium density profile is expanded in discrete modes
\begin{align}
 n(z)=\frac{1}{L}\sum_{\mu}n_{\mu}\exp{(-\imag Q_{\mu}z)},
\end{align}
with corresponding Fourier coefficients
\begin{align}
 n_{\mu}=\int_{-L/2}^{L/2}\mathrm{d}z n(z)\exp{(\imag Q_{\mu}z)}.
\end{align}
Since $n(z)$ is real, the Fourier coefficients fulfil $n_{\mu}=n_{-\mu}^*$. For the case under consideration, the density profile is also symmetric $n(z) = n(-z)$ which implies $n_\mu \in \mathbb{R}$ is real. We shall also use the local specific volume  defined by $v(z)\coloneqq1/n(z)$.
Using the convolution theorem the Fourier coefficients fulfil
\begin{align}
 \sum_{\kappa}n_{\mu-\kappa}v_{\kappa-\nu}=\sum_{\kappa}n_{\mu-\kappa}^*v_{\kappa-\nu}^*=L^2\delta_{\mu\nu}.
\end{align}
The mode decomposition of the microscopic density is now
\begin{equation}
 \rho(\vec{r},z)=\frac{1}{A}\sum_{\vec{q}}\frac{1}{L}\sum_\mu \rho_\mu(\vec{q})
  \exp{\left(-\imag Q_\mu z\right)}e^{-\imag\vec{q}\cdot\vec{r}},
\end{equation}
where  the  expansion coefficients
\begin{equation}
 \rho_\mu(\vec{q}) = \sum_{n=1}^N \exp{\left[\imag Q_\mu z_n\right]}
  e^{\imag \vec{q}\cdot\vec{r}_n},
\end{equation}
are treated as the fundamental quantities. We shall be interested in dynamic correlation functions of  the fluctuations $\delta \rho_\mu(\vec{q}) \coloneqq \rho_\mu(\vec{q})-\langle \rho_\mu(\vec{q}) \rangle$. However, since 
\begin{equation}
 \langle \rho_\mu(\vec{q}) \rangle = A n_\mu \delta_{\vec{q},0},
\end{equation}
the distinction is relevant only for vanishing wave vector $\vec{q}=0$ parallel to the plates. 

Then the corresponding mode expansion of the Van Hove function
\begin{align}
 G(|\vec{r}-\vec{r}^{\,\prime}|,z,z^\prime,t) =& \frac{1}{A}\sum_{\vec{q}}\frac{1}{L^2}\sum_{\mu\nu}
  S_{\mu\nu}(\vec{q},t) e^{\imag\vec{q}\cdot\left(\vec{r}-\vec{r}^{\,\prime}\right)}\exp{\left[\imag \left(Q_\mu z - Q_\nu z^\prime\right)\right]},
\end{align}
is achieved in terms of the  generalized intermediate scattering function
\begin{equation}
 S_{\mu\nu}(q,t)=\frac{1}{N}\langle \delta\rho_\mu(\vec{q},t)^*\delta \rho_\nu(\vec{q})\rangle.
\end{equation}
Due to translational invariance parallel to the plates, only correlation functions for the same  wave vector $\vec{q}$ are non-vanishing, while rotational invariance entails that the intermediate scattering function depends only on the magnitude of the wave vector $q = |\vec{q}|$. 

Reversely, the intermediate scattering function is obtained by taking the corresponding Fourier integrals 
\begin{align}
 S_{\mu\nu}(q,t) =& \int_{-L/2}^{L/2}\mathrm{d}z \int_{-L/2}^{L/2}\mathrm{d}z^\prime \int_A \mathrm{d}(\vec{r}-\vec{r}^{\,\prime})
  G(|\vec{r}-\vec{r}^{\,\prime}|,z,z^\prime,t) \nonumber\\
  &\times\exp{\left[-\imag \left( Q_\mu z - Q_\nu z^\prime\right)\right]}e^{-\imag\vec{q}\cdot\left(\vec{r}-\vec{r}^{\,\prime}\right)} . 
\end{align}
The initial value $S_{\mu\nu}(q)\coloneqq S_{\mu\nu}(q,t=0)$ characterizing the equilibrium structure of the fluid takes the role of a generalized static structure factor for the slit geometry. 

For each wavenumber $q$ the matrix  $S_{\mu\nu}(q,t)$ fulfils the properties of a matrix-valued correlation function and for the Brownian dynamics considered here also the ones of a matrix-valued completely monotone function. Hence, the contractions $\sum_{\mu \nu} y_\mu^* S_{\mu\nu}(q,t) y_\nu$ may be represented as a characteristic function of a finite Lebesgue-Stieltjes measure and for our case also as a Laplace transform of another Lebesgue-Stieltjes measure as in Eq.~\eqref{eq:Bernstein}.

\section{Zwanzig-Mori projection operator formalism for Brownian dynamics}
In this section we derive exact equations of motion (e.o.m.) for the generalized intermediate scattering function $S_{\mu\nu}(q,t)$ relying on the Zwanzig-Mori procedure.

For a pair of  orthogonal projection operators $\mathcal{P}=1-\mathcal{Q}$ the following operator identity holds~\cite{Hansen:Theory_of_Simple_Liquids,Lang:PRE:2012}
\begin{align}
\mathcal{P}\partial_t \mathcal{R}(t)\mathcal{P} =& \mathcal{P}\Omega^\dagger\mathcal{P}
  \mathcal{R}(t)\mathcal{P} +\int_0^t\mathcal{P}\Omega^\dagger\mathcal{Q}e^{\mathcal{Q}\Omega^\dagger\mathcal{Q}(t-t^\prime)}
  \mathcal{Q}\Omega^\dagger\mathcal{P}\mathcal{R}(t^\prime)\mathcal{P}\mathrm{d}t^\prime.
\end{align}
To derive the e.o.m.\ for the generalized intermediate scattering function $S_{\mu\nu}(q,t)$ the projector
\begin{equation}\label{eq:projector}
\mathcal{P}_\rho=\frac{1}{N}\sum_{\vec{q}}\sum_{\mu\nu}\ket{\rho_\mu(\vec{q})}\left[\mathbf{S}^{-1}(q)\right]_{\mu\nu}
  \bra{\rho_\nu(\vec{q})},
\end{equation}
is introduced, where the density modes are used as distinguished variables. Then the  exact first e.o.m.\
\begin{align}
 \dot{S}_{\mu\nu}(q,t) &+ \sum_{\kappa\lambda}D_{\mu\kappa}(q)[\mathbf{S}^{-1}(q)]_{\kappa\lambda}S_{\lambda\nu}(q,t) \nonumber\\
  &+ \sum_{\kappa\lambda}\int_0^t\delta K_{\mu\kappa}(q,t-t^\prime) [\mathbf{S}^{-1}(q)]_{\kappa\lambda} S_{\lambda\nu}(q,t^\prime)
  \mathrm{d}t^\prime = 0,
\end{align}
follows. Here, the short-time diffusion coefficient reads
\begin{equation}
 D_{\mu\nu}(q) = -\frac{1}{N}\braket{\rho_\mu(\vec{q})|\Omega^\dagger\rho_\nu(\vec{q})},
\end{equation}
and the memory kernel is expressed as
\begin{equation}
 \delta K_{\mu\nu}(q,t)=-\frac{1}{N}\braket{\mathcal{Q}_\rho\Omega^\dagger\rho_\mu(\vec{q})|
  e^{\mathcal{Q}_\rho\Omega^\dagger\mathcal{Q}_\rho t}|\mathcal{Q}_\rho\Omega^\dagger\rho_\nu(\vec{q})}.
\end{equation}
Since the operator $-\mathcal{Q}_\rho \Omega^\dagger \mathcal{Q}_\rho$ is again hermitian with respect to the Kubo scalar product and displays a non-negative spectrum, for each wavenumber $q$ the time-dependent matrix $-\delta K_{\mu\nu}(q,t)$ corresponds to a matrix-valued completely monotone function.

For future manipulations we rewrite the e.o.m.\ in obvious  matrix representation
\begin{align}
 \dot{\mathbf{S}}(q,t)+\mathbf{D}(q)\mathbf{S}^{-1}(q)\mathbf{S}(q,t) 
  +\int_0^t\delta\mathbf{K}(q,t-t^\prime)\mathbf{S}^{-1}(q)\mathbf{S}(q,t^\prime)
  \mathrm{d}t^\prime=0.
\end{align}
This e.o.m.\ has been anticipated earlier in Ref.~\cite{Lang:JStatMech:2013}, here we provide the microscopic derivation. Furthermore microscopic expressions for the memory kernel and the diffusion matrix are formulated. 

The explicit expression for the latter one is derived in appendix \ref{appendix_diffusion},
\begin{equation}
 D_{\mu\nu}(q)=D_0 \frac{n_{\mu-\nu}^*}{n_0} (q^2+Q_\mu Q_\nu).
\end{equation}
We also calculate the 'fluctuating force' with selector 
$b^\alpha(x,z)= x\delta_{\alpha,\parallel}+z\delta_{\alpha,\perp}$ explicitly (appendix \ref{appendix_fluctuatingforce})
\begin{equation}
 \mathcal{Q}_{\rho}\Omega^\dagger\ket{\rho_\mu(\vec{q})} = 
\imag\frac{D_0}{k_B T} \sum_{\alpha}b^\alpha(q,Q_{\mu}) \mathcal{Q}_\rho \ket{F^{\alpha}_{\mu}(\vec{q})},
\end{equation}
where we defined the fluctuating force 
\begin{align}\label{eq:fluctuating_forces}
 \delta F_{\mu}^\alpha(\vec{q}) = - \sum_{n=1}^N  b^\alpha\left(\frac{\partial U}{\partial \vec{r}_n}, \frac{\partial U}{\partial z_n}\right) e^{\imag\vec{q}\cdot\vec{r}_n}
  \exp{(\imag Q_{\mu}z_n)} + \imag k_B T b^\alpha(q,Q_\mu)\braket{\rho_{\mu}(\vec{q})},
\end{align}
for the two relaxation channels $\alpha= \parallel, \perp$.
The memory kernel thus splits naturally into two relaxation channels
\begin{equation}
 \delta K_{\mu\nu}(q,t) = \sum_{\alpha\beta=\parallel,\perp}b^\alpha(q,Q_\mu)\delta \mathcal{K}_{\mu\nu}^{\alpha\beta}(q,t)
  b^\beta(q,Q_\nu),
\end{equation}
with reduced memory kernel 
\begin{align}
 \delta\mathcal{K}_{\mu\nu}^{\alpha\beta}(q,t) =-\left(\frac{D_0}{k_B T}\right)^2 \frac{1}{N}
  \braket{F_\mu^\alpha(\vec{q})|\mathcal{Q}_{\rho}e^{\mathcal{Q}_\rho\Omega^\dagger\mathcal{Q}_\rho t}\mathcal{Q}_{\rho}
  |F_\nu^\beta(\vec{q})}.
\end{align}
Again $-\delta \mathcal{K}_{\mu\nu}^{\alpha\beta}(q,t)$ is a matrix-valued completely monotone function. Here the matrix has to be read  with respect to the super-index $(\alpha, \mu)$ consisting of the mode-index $\mu$ and the channel index $\alpha$. 

We note that the matrix of short-time diffusion coefficients displays the same natural splitting
\begin{align}
  D_{\mu\nu}(q)
  &= \sum_{\alpha\beta=\parallel,\perp}b^{\alpha}(q,Q_\mu)\mathcal{D}_{\mu\nu}^{\alpha\beta}(q)b^{\beta}(q,Q_\nu),
\end{align}
with the channel diffusion matrix
\begin{equation}
 \mathcal{D}_{\mu\nu}^{\alpha\beta}(q) = D_0\frac{n_{\mu-\nu}^*}{n_0}\delta_{\alpha\beta}.
\end{equation}

\section{Irreducible memory kernel}
In this section we derive exact e.o.m.\ for the matrix-valued kernel $\delta \mathcal{K}_{\mu\nu}^{\alpha\beta}(q,t)$ in terms of an irreducible memory kernel. For the bulk case, where no splitting of the relaxation channels occurs, this has been elaborated by Cichocki and Hess~\cite{Cichocki:PhysicaA:1987} and later more generally by Kawasaki~\cite{Kawasaki:PhysicaA:1995}. The derivation for confined liquids presented here is new and extends the previous approaches to multi-channel relaxation. For the case of microrheology, where relaxation parallel and perpendicular to the external force emerge, irreducible memory functions have already been used successfully in MCT approaches~\cite{Gruber:PRE:2016,Gruber:PHD:2019}.

The key insight is that the  operator identity
\begin{equation}\label{eq:operator_id}
 e^{\mathcal{Q}_\rho\Omega^\dagger\mathcal{Q}_\rho t}=e^{\Omega_{\text{irr}}^\dagger t}+\int_0^t 
  e^{\Omega_{\text{irr}}^\dagger(t-t^\prime)}\delta\Omega e^{\mathcal{Q}_\rho\Omega^\dagger\mathcal{Q}_\rho t^\prime}
  \mathrm{d}t^\prime,
\end{equation}
is valid for arbitrary splitting $\mathcal{Q}_\rho \Omega^\dagger\mathcal{Q}_\rho =\Omega_{\text{irr}}^\dagger+\delta\Omega$. Therefore exact e.o.m.\ of the Zwanzig-Mori type can be derived within any splitting scheme. Yet, a proper  choice of the irreducible (adjoint) Smoluchowski operator $\Omega_{\text{irr}}^\dagger$ is crucial to make further analytic progress. In particular, we require that the splitting suggests a mode-coupling approximation with a mode-coupling functional that is identical to the case of Newtonian dynamics. Here we propose to define the irreducible operator by 
\begin{align}
  \Omega_{\text{irr}}^\dagger :=  \mathcal{Q}_\rho \Omega^\dagger\mathcal{Q}_\rho + \frac{1}{N}\left(\frac{D_0}{k_B T}\right)^2  \sum_{\alpha\beta=\parallel,\perp} \sum_{\mu\nu} \mathcal{Q}_{\rho}\ket{F_{\mu}^{\alpha}(\vec{q})} 
  \left[\bm{\mathcal{D}}^{-1}(q)\right]_{\mu\nu}^{\alpha\beta} \bra{F_{\nu}^{\beta}(\vec{q})}\mathcal{Q}_{\rho} .
\end{align}
Sandwiching  the operator identity, Eq.~\eqref{eq:operator_id}, between the fluctuating forces, Eq.~\eqref{eq:fluctuating_forces}, projected onto the orthogonal subspace, the second exact e.o.m.\ is found
\begin{align}\label{eq:second_eom}
 \delta\mathcal{K}_{\mu\nu}^{\alpha\beta}(q,t)=  -\mathfrak{M}^{\alpha\beta}_{\mu\nu}(q,t) - \sum_{\kappa\lambda}
  \sum_{\gamma=\parallel,\perp}\int_0^t \mathfrak{M}^{\alpha\gamma}_{\mu\kappa}(q,t-t^\prime) 
  \left[\bm{\mathcal{D}}^{-1}(q)\right]_{\kappa\lambda}^{\gamma\gamma}\delta\mathcal{K}_{\lambda\nu}^{\gamma\beta}(q,t^\prime)
  \mathrm{d}t^\prime,
\end{align}
with the irreducible memory kernel
\begin{equation}
\mathfrak{M}^{\alpha\beta}_{\mu\nu}(q,t) = \frac{1}{N}\left(\frac{D_0}{k_B T}\right)^2
  \braket{F_{\mu}^{\alpha}(\vec{q})|\mathcal{Q}_{\rho}e^{\Omega_{\text{irr}}^\dagger t}\mathcal{Q}_{\rho}|
  F_{\nu}^{\beta}(\vec{q})}.
\end{equation}
It is favourable to introduce the  effective force kernel $\bm{\mathcal{M}}(q,t):=\bm{\mathcal{D}}^{-1}(q)\bm{\mathfrak{M}}(q,t)\bm{\mathcal{D}}^{-1}(q)$ by stripping off a channel diffusion matrix $\bm{\mathcal{D}}$ from the left and right. Here the bold calligraphic symbols are the   matrix notation with matrix elements specified by a column and row superindex: $ [\bm{\mathcal{M}}(q,t)]^{\alpha \beta}_{\mu\nu} = \mathcal{M}^{\alpha\beta}_{\mu\nu}(q,t)$. Then the second e.o.m.~\eqref{eq:second_eom} can be rearranged,
\begin{align}
 \delta\bm{\mathcal{K}}(q,t)=-\bm{\mathcal{D}}(q)\bm{\mathcal{M}}(q,t)\bm{\mathcal{D}}(q)
  -\int_0^t \bm{\mathcal{D}}(q)\bm{\mathcal{M}}(q,t-t^\prime)\delta\bm{\mathcal{K}}(q,t^\prime) \mathrm{d}t^\prime,
\end{align}
consistent with e.o.m.\ Eq.~(23) in Ref.~\cite{Lang:JStatMech:2013}. Due to the multiple decay channels both integro-differential equations cannot be combined to one single integro-differential equation (like in the case of simple liquids), just as in the case of Newtonian dynamics~\cite{Lang:PRE:2012}.

Introducing the Fourier-Laplace domain convention,
\begin{align}
 \hat{S}_{\mu\nu}(q,z)=\imag\int_0^\infty S_{\mu\nu}(q,t)\exp(\imag z t)\mathrm{d}t, \quad \text{Im}[z]>0,
\end{align}
with complex frequency $z$ the e.o.m.\ can be simplified. Here the Laplace transforms are analytic functions in the upper half plane $\mathbb{C}_{+}=\{z\in\mathbb{C}|\text{Im}[z]>0\}$. The decomposition into multiple relaxation channels is directly converted to the Fourier-Laplace domain due to linearity.

We define a modified memory kernel $\bm{\mathcal{\hat{K}}}(q,z)=\delta\bm{\mathcal{\hat{K}}}(q,z)+\imag\bm{\mathcal{D}}(q)$, which explicitly takes the high-frequency limit $\bm{\mathcal{\hat{K}}}(q,z)\to\imag\bm{\mathcal{D}}(q)$ as $z\to\infty$ into account and implies a $\delta$-function at the time origin.

The contraction is split accordingly, $\mathbf{\hat{K}}(q,z)=\delta\mathbf{\hat{K}}(q,z)+\imag\mathbf{D}(q)$. Then the transformed first equation of motion yields a matrix equation with formal solution in the frequency domain~\cite{Lang:JStatMech:2013}
\begin{align}
 \mathbf{\hat{S}}(q,z)=-\Big[z\mathbf{S}^{-1}(q)+\mathbf{S}^{-1}(q)\mathbf{\hat{K}}(q,z)\mathbf{S}^{-1}(q)\Big]^{-1},
\end{align}
and from the second equation the memory kernel
\begin{align}
 \bm{\mathcal{\hat{K}}}(q,z)=-\Big[\imag\bm{\mathcal{D}}^{-1}(q)+\bm{\mathcal{\hat{M}}}(q,z)\Big]^{-1},
\end{align}
can be calculated. So far, the e.o.m.\ are exact and all interactions (within the liquid as well as with the confinement) are hidden in the irreducible memory kernel $\bm{\mathcal{\hat{M}}}(q,z)$.

\section{Mode-coupling theory}
The density dynamics are expressed in terms of the irreducible memory kernel
\begin{equation}
 \mathfrak{M}^{\alpha\beta}_{\mu\nu}(q,t) =\frac{1}{N}\left(\frac{D_0}{k_B T}\right)^2\braket{F_\mu^\alpha(\vec{q})|
\mathcal{Q}_{\rho}e^{\Omega_{\text{irr}}^\dagger t}\mathcal{Q}_{\rho}|F_\nu^\beta(\vec{q})},
\end{equation}
within the Zwanzig-Mori formalism. A self-consistent solution of the two coupled e.o.m.\ relies on convenient mode-coupling approximations. Due to the caging by neighbouring particles the dynamics are slowed down. These caging forces entering the memory kernel originate from the interactions with the particles (therefore by products of density modes). Following the mode-coupling theory for supercooled liquids~\cite{Goetze:Complex_Dynamics} and confined Newtonian liquids~\cite{Lang:PRE:2012} an expression for the irreducible memory kernel as a functional of the generalized intermediate scattering function is provided. 

We start with the well-established idea of projecting the forces onto a set of fluctuating density-pair modes $\mathcal{Q}_{\rho}\ket{F_\mu^\alpha(\vec{q})}\approx \mathcal{P}_{\rho\rho}\mathcal{Q}_{\rho}\ket{F_\mu^\alpha(\vec{q})}$, where the projection operator onto the pair fluctuating modes reads~\cite{Lang:PRE:2012}
\begin{align}
 \mathcal{P}_{\rho\rho}=&\sum_{\substack{\mu_1,\mu_2\\\mu_1^\prime,\mu_2^\prime}}
  \sum_{\substack{\vec{q}_1,\vec{q}_2\\ \vec{q}_1^{\,\prime},\vec{q}_2^{\,\prime}}}\ket{\rho_{\mu_1}(\vec{q}_1)\rho_{\mu_2}(\vec{q}_2)} g_{\mu_1\mu_2;\mu_1^\prime\mu_2^\prime}
  (\vec{q}_1\vec{q}_2,\vec{q}_1^{\,\prime} \vec{q}_2^{\,\prime}) \bra{\rho_{\mu_1^\prime}(\vec{q}_1^{\,\prime})\rho_{\mu_2^\prime}(\vec{q}_2^{\,\prime})}.
\end{align}
From now on we use a simplified notation following Ref.~\cite{Scheidsteger:PRE:1997} by using superindices $i=(\vec{q}_i,\mu_i)$, where wave vectors and mode indices are coupled together. To ensure idempotency ($\mathcal{P}_{\rho\rho}^2=\mathcal{P}_{\rho\rho}$) the normalization condition
\begin{align}
 \sum_{1^\prime 2^\prime} g(12;1^\prime 2^\prime)\braket{\delta\rho(1^\prime)^*\delta\rho(2^\prime)^*
  \delta\rho(1^{\prime\prime})\delta\rho(2^{\prime\prime})} = \frac{1}{2}\left[\delta(1,1^{\prime\prime})\delta(2,2^{\prime\prime})+\delta(1,2^{\prime\prime})\delta(2,1^{\prime\prime})\right],
\end{align}
for the matrix $g(12;1^\prime 2^\prime)$ has to be fulfilled~\cite{Lang:PRE:2012}. The crucial step of mode-coupling theory is to factorize the dynamical four-point correlation function into dynamical two-point correlation functions~\cite{Lang:PRE:2012}
\begin{align}
 \langle\delta\rho(1)^*\delta\rho(2)^*\exp{\left(\Omega_{\text{irr}}^\dagger t
  \right)}\delta\rho(1^\prime)\delta\rho(2^\prime)\rangle  \approx N^2 \left[S(1,1^\prime,t)S(2,2^\prime,t) + S(1,2^\prime,t)S(2,1^\prime,t)\right],
\end{align}
and, in particular, for the static correlation function at the time origin
\begin{align}
 \langle\delta\rho(1)^*\delta\rho(2)^*\delta\rho(1^\prime)\delta\rho(2^\prime)\rangle \approx \braket{\rho(1)|\rho(1^\prime)}\braket{\rho(2)|\rho(2^\prime)} +
  \braket{\rho(1)|\rho(2^\prime)}\braket{\rho(2)|\rho(1^\prime)}.
\end{align}
The same factorization allows us to determine consistently the approximate normalization matrix~\cite{Lang:PRE:2012}
\begin{align}
g(12,1^\prime 2^\prime) \approx
  \frac{1}{4N^2}\left\{\left[\mathbf{S}^{-1}\right](1,1^\prime)\left[\mathbf{S}^{-1}\right](2,2^\prime)
  + \left[\mathbf{S}^{-1}\right](1,2^\prime)\left[\mathbf{S}^{-1}\right](2,1^\prime)\right\}.
\end{align}
Aggregating the terms together the memory kernel is approximated by a bilinear functional of the generalized intermediate scattering function~\cite{Lang:PRE:2012}
\begin{align}
  \mathfrak{M}^{\alpha\beta}_{\mu\nu}(q,t)\approx & \frac{1}{2N^3}\left(\frac{D_0}{k_B T}\right)^2 \sum_{\substack{\vec{q}_1 \\ \vec{q}_2=\vec{q}-\vec{q}_1}}\sum_{\substack{\mu_1\mu_2\\ \nu_1\nu_2}} 
  \mathcal{X}^{\alpha}_{\mu,\mu_{1}\mu_{2}}(\vec{q},\vec{q}_{1}\vec{q}_{2})
  \nonumber \\
  & \times S_{\mu_1\nu_1}(q_1,t)S_{\mu_2\nu_2}(q_2,t)\mathcal{X}^{\beta}_{\nu,\nu_{1}\nu_{2}}(\vec{q},\vec{q}_{1}\vec{q}_{2})^*.
\end{align}
Due to translational invariance in lateral direction to the walls only wave vectors $\vec{q}_1$, $\vec{q}_2$ with the selection rule $\vec{q}=\vec{q}_1+\vec{q}_2$ contribute. The complex-valued vertex $\mathcal{X}^{\alpha}_{\mu,\mu_{1}\mu_{2}}(\vec{q},\vec{q}_{1}\vec{q}_{2})$ is generated from the overlap of the fluctuating forces with the density pair modes
\begin{align}
\mathcal{X}^{\alpha}_{\mu,\mu_{1}\mu_{2}}(\vec{q},\vec{q}_{1}\vec{q}_{2}) = & \sum_{\mu_1^\prime\mu_2^\prime}
  \braket{F_\mu^\alpha(\vec{q})|\mathcal{Q}_{\rho}\rho_{\mu_1^\prime}(\vec{q}_1)\rho_{\mu_2^\prime}(\vec{q}_2)} \left[\mathbf{S}^{-1}(q_1)\right]_{\mu_1^\prime\mu_1}\left[\mathbf{S}^{-1}(q_2)\right]_{\mu_2^\prime\mu_2}.
\end{align}
The main difference in changing the dynamics from Newtonian~\cite{Lang:PRE:2012} to Brownian is the calculation of the overlap matrix elements in terms of structural quantities (see appendix \ref{appendix_overlap}).
Introducing a triple correlation function
\begin{equation}
 S_{\sigma,\mu_1\mu_2}(\vec{q},\vec{q}_1\vec{q}_2) = \frac{1}{N}\braket{
  \delta\rho_{\sigma}(\vec{q})^*\delta\rho_{\mu_1}(\vec{q}_1)\delta\rho_{\mu_2}(\vec{q}_2)},
\end{equation}
the resulting explicit expression is identical to the Newtonian case
\begin{align}\label{eq:overlap_explicit}
\langle F_{\mu}^{\alpha}(\vec{q})|\mathcal{Q}_{\rho}\rho_{\mu_1}(\vec{q}_1)\rho_{\mu_2}(\vec{q}_2)\rangle = &- \imag N k_B T \delta_{\vec{q},\vec{q}_1+\vec{q}_2} \Big[
b^{\alpha}(\hat{\vec{q}}\cdot\vec{q}_1,Q_{\mu_1})
  S_{\mu-\mu_1,\mu_2}(q_2) + (1 \leftrightarrow 2) \nonumber\\
 &-\sum_{\kappa\sigma} \frac{n_{\mu-\kappa}^*}{n_0} b^{\alpha}(q,Q_\kappa)\left[\mathbf{S}^{-1}(q)\right]_{\kappa\sigma} S_{\sigma,\mu_1\mu_2}(\vec{q},\vec{q}_1\vec{q}_2)\Big],
\end{align}
provided one identifies the fluctuating force $F_{\mu}^{\alpha}(\vec{q})$  with time derivatives of the channel-resolved fluctuating momentum fluxes~\cite{Lang:PRE:2012}.
This remarkable property is one of the main results of this work and corroborates the MCT approach to describe the slow structural relaxation. Physically it reflects that only the internal stresses between the colloidal particles lead to glassy dynamics, which are balanced by Newton's third law, while the interaction with the solvent merely changes the short-time dynamics.

Since the triple correlation functions in Eq.~\eqref{eq:overlap_explicit} are difficult to evaluate further approximations are adopted. Glassy dynamics are successfully described by using the convolution approximation~\cite{Goetze:Complex_Dynamics}, which has been effectively expanded to a liquid confined in a slit~\cite{Lang:PRE:2012}. Since the convolution approximation does not depend on the microscopic dynamics it is not repeated here. Expressing the static three-point correlation function in terms of products of two-point correlation functions the vertices take the compact form
\begin{align}
 &\mathcal{X}^{\alpha}_{\mu,\mu_{1}\mu_{2}}(\vec{q},\vec{q}_{1}\vec{q}_{2})\approx \imag N k_B T \frac{n_0}{L^2}
  \delta_{\vec{q},\vec{q}_1+\vec{q}_2}\Big[b^{\alpha}(\hat{\vec{q}}\cdot\vec{q}_1,Q_{\mu-\mu_2})c_{\mu-\mu_2,\mu_1}(q_1) + (1 \leftrightarrow 2)\Big].
\end{align}
The matrix elements $c_{\mu\nu}(q)$ of the direct correlation function are defined by a generalized Ornstein-Zernike equation~\cite{Lang:PRE:2012}
\begin{equation}\label{eq:Ornstein-Zernike}
 \mathbf{S}^{-1}(q) = \frac{n_0}{L^2}\left[\mathbf{v}-\mathbf{c}(q)\right],
\end{equation}
with $\left[\mathbf{v}\right]_{\mu\nu}=v_{\nu-\mu}$.

The effective force kernel is then given by
\begin{align}
 \mathcal{M}_{\mu\nu}^{\alpha\beta}(q,t) =& \left[\bm{\mathcal{D}}^{-1}(q)\bm{\mathfrak{M}}(q,t)\bm{\mathcal{D}}^{-1}(q)\right]
  _{\mu\nu}^{\alpha\beta} \nonumber \\ \approx& 
  \frac{1}{2N}\sum_{\substack{\vec{q}_1\\\vec{q}_2=\vec{q}-\vec{q}_1}}
  \sum_{\substack{\mu_1\mu_2\\\nu_1\nu_2}} \mathcal{Y}_{\mu,\mu_1\mu_2}^\alpha
   (\vec{q},\vec{q}_1\vec{q}_2)S_{\mu_1\nu_1}(q_1,t) S_{\mu_2\nu_2}(q_2,t) \mathcal{Y}_{\nu,\nu_1\nu_2}^\beta
  (\vec{q},\vec{q}_1\vec{q}_2)^*,
\end{align}
with new vertices
\begin{align}
 \mathcal{Y}_{\mu,\mu_1\mu_2}^\alpha & (\vec{q},\vec{q}_1\vec{q}_2) = \frac{n_0^2}{L^4}\delta_{\vec{q},\vec{q}_1+\vec{q}_2}
  \sum_{\kappa}v_{\mu-\kappa}^* \Big[b^\alpha(\hat{\vec{q}}\cdot\vec{q}_1,Q_{\kappa-\mu_2})
  c_{\kappa-\mu_2,\mu_1}(q_1) + (1 \leftrightarrow 2) \Big].
\end{align}
The bilinear form $\mathcal{M}_{\mu\nu}^{\alpha\beta}(q,t)=\mathcal{F}_{\mu\nu}^{\alpha\beta}[\mathbf{S}(t),\mathbf{S}(t);q]$ of the force kernel is a direct consequence of the mode-coupling approximations.
In the long-wavelength limit the force kernel decouples  with respect to the channel index $\mathcal{M}_{\mu\nu}^{\alpha\beta}(q\to 0,t)\eqqcolon \delta^{\alpha\beta}\mathcal{M}_{\mu\nu}^{\alpha}(t)$ and the vertex vanishes as $O(q)$ just as in the momentum-conserving case~\cite{Lang:PRE:2012}.

\section{Tagged-particle motion}
It is now straightforward to derive also the e.o.m.\ and the MCT functional for the case of a tagged-particle with arbitrary size 
(of particular interest since self-dynamics are directly accessible in computer simulations and single-particle tracking experiments) by adapting the strategy elaborated above. Rather than repeating all the steps we provide  a short summary of the basic equations for future reference. 

Since the tracer particle may differ from the solvent particles, its accessible slit width is denoted by $L_s\ne L$. The incoherent intermediate scattering function reads
\begin{equation}
 S_{\mu\nu}^{(s)}(q,t)=\langle \delta\rho_\mu^{(s)}(\vec{q},t)^*\delta \rho_\nu^{(s)}(\vec{q})\rangle,
\end{equation}
where
\begin{equation}
 \rho_\mu^{(s)}(\vec{q}) = \exp{\left[\imag Q_\mu^{(s)} z_s\right]}e^{\imag \vec{q}\cdot\vec{r}_s},
\end{equation}
are the density modes ($Q_\mu^{(s)}=2\pi\mu/L_s, \mu\in\mathbb{Z}$) of the microscopic density $\rho^{(s)}(\vec{r},z)$. The initial value $S_{\mu\nu}^{(s)}(q,t=0)\eqqcolon S_{\mu\nu}^{(s)}=n_{\mu-\nu}^{(s)*}/n_0$ is solely identified by the density modes. Using Zwanzig-Mori technique with a suitable tagged-particle projector~\cite{Lang:PRE_89:2014}
\begin{equation}\label{eq:projector_tagged}
\mathcal{P}_{\rho^{(s)}}=\sum_{\mu\nu}\ket{\rho_\mu^{(s)}(\vec{q})}\left[(\mathbf{S}^{(s)}){}^{-1}\right]_{\mu\nu}\bra{\rho_\nu^{(s)}(\vec{q})},
\end{equation}
the exact first e.o.m.\ in matrix representation reads
\begin{align}
 \dot{\mathbf{S}}^{(s)}(q,t)+\mathbf{D}^{(s)}(q)(\mathbf{S}^{(s)}){}^{-1}
\mathbf{S}^{(s)}(q,t)+\int_0^t\delta\mathbf{K}^{(s)}(q,t-t^\prime)
(\mathbf{S}^{(s)}){}^{-1}
\mathbf{S}^{(s)}(q,t^\prime)\mathrm{d}t^\prime=0.
\end{align}
The short-time self-diffusion coefficient can be  expressed explicitly by
\begin{equation}
 D_{\mu\nu}^{(s)}(q) = D_0 \frac{n_{\mu-\nu}^{(s)*}}{n_0} (q^2+Q_\mu^{(s)} Q_\nu^{(s)}),
\end{equation}
and the self-memory kernel $\delta K_{\mu\nu}^{(s)}$ including the history of the dynamical process naturally splits into two relaxation channels
\begin{equation}
 \delta K_{\mu\nu}^{(s)}(q,t) = \sum_{\alpha\beta=\parallel,\perp}b^\alpha(q,Q_\mu^{(s)})\delta \mathcal{K}_{\mu\nu}^{\alpha\beta,(s)}(q,t)
  b^\beta(q,Q_\nu^{(s)}).
\end{equation}
The crucial next step is to identify a suitable irreducible operator for the self-dynamics such that the splitting into multiple relaxation channels is properly reflected. We employ the following choice: 
\begin{align}
  \Omega_{\text{irr}}^{\dagger,(s)} :=  \mathcal{Q}_{\rho^{(s)}} \Omega^\dagger\mathcal{Q}_{\rho^{(s)}} &+ \left(\frac{D_0}{k_B T}\right)^2  \sum_{\alpha\beta=\parallel,\perp} \sum_{\mu\nu} \mathcal{Q}_{{\rho^{(s)}}}\ket{F_{\mu}^{\alpha,(s)}(\vec{q})} \nonumber \\ &\times\left[(\bm{\mathcal{D}}^{(s)}(q)){}^{-1}\right]_{\mu\nu}^{\alpha\beta} \bra{F_{\nu}^{\beta,(s)}(\vec{q})}\mathcal{Q}_{\rho^{(s)}}.
\end{align}
The second e.o.m.\ for the reduced self-memory kernel $\delta\bm{\mathcal{K}}^{(s)}(q,t)$ is then found to be
\begin{align}
 \delta\bm{\mathcal{K}}^{(s)}(q,t)=&-\bm{\mathcal{D}}^{(s)}(q)\bm{\mathcal{M}}^{(s)}(q,t)\bm{\mathcal{D}}^{(s)}(q) -\int_0^t \bm{\mathcal{D}}^{(s)}(q)\bm{\mathcal{M}}^{(s)}(q,t-t^\prime)\delta\bm{\mathcal{K}}^{(s)}(q,t^\prime)\mathrm{d}t^\prime,
\end{align}
where also the channel diffusion matrix $\mathcal{D}_{\mu\nu}^{\alpha\beta,(s)}(q) := D_0\delta_{\alpha\beta} n_{\mu-\nu}^{(s)*}/n_0$ enters.

Using MCT for a suitable description of the local caging forces, the effective force kernel
\begin{align}
 \mathcal{M}_{\mu\nu}^{\alpha\beta,(s)}(q,t) \approx
  \frac{1}{N}\sum_{\substack{\vec{q}_1\\\vec{q}_2=\vec{q}-\vec{q}_1}}
  \sum_{\substack{\mu_1\mu_2\\\nu_1\nu_2}} \mathcal{Y}_{\mu,\mu_1\mu_2}^{\alpha,(s)}
   (\vec{q},\vec{q}_1\vec{q}_2)S_{\mu_1\nu_1}(q_1,t) S_{\mu_2\nu_2}^{(s)}(q_2,t) \mathcal{Y}_{\nu,\nu_1\nu_2}^{\beta,(s)}
  (\vec{q},\vec{q}_1\vec{q}_2)^*,
\end{align}
is a linear functional of the incoherent scattering function of the tracer and the collective intermediate scattering function of the host liquid. The vertices are given by
\begin{align}
 \mathcal{Y}_{\mu,\mu_1\mu_2}^{\alpha,(s)} (\vec{q},\vec{q}_1\vec{q}_2) = \frac{n_0}{L L_s}\delta_{\vec{q},\vec{q}_1+\vec{q}_2}
  \sum_{\sigma}\left[(\mathbf{S}^{(s)}){}^{-1}\right]_{\mu\sigma} 
b^\alpha(\hat{\vec{q}}\cdot\vec{q}_1,Q_{\sigma-\mu_2}^{(s)})
  c_{\sigma-\mu_2,\mu_1}^{(s)}(q_1),
\end{align}
with the generalized direct correlation function $c_{\mu\nu}^{(s)}(q)$ determining the static interactions between the single tagged-particle and the complex environment of the host liquid. It is related to the overlap of the tagged-particle density and the collective density by a generalized Ornstein-Zernike equation~\cite{Lang:PRE_89:2014}. In contrast to the collective dynamics no convolution approximation is required since the vertex does not contain any triple correlation function. In the long-wavelength limit $q\to 0$ the functional $\mathcal{M}_{\mu\nu}^{\alpha\beta,(s)}(q\to 0,t)\eqqcolon\delta^{\alpha\beta}\mathcal{M}_{\mu\nu}^{\alpha}(t)$ decouples as in the collective case, but in contrast to collective dynamics the vertex does not vanish for $q\to 0$ since the corresponding vertex does not depend explicitly on $q$ anymore.

Again, the resulting expression is identical to the case of Newtonian dynamics, if the fluctuating force is replaced by time derivatives of the channel-resolved fluctuating momentum fluxes~\cite{Lang:PRE_89:2014}.

\section{Summary and conclusions}
In this work a mode-coupling theory of the glass transition for a colloidal suspension confined in a slit geometry has been derived microscopically. The theory yields  a closed set of equations to be solved self-consistently, both for the collective and the tagged-particle dynamics. Although the e.o.m.\ have been anticipated  before~\cite{Lang:JStatMech:2013} by matching the short-time dynamics to the mode-coupling functional, a  microscopic approach has been missing, a gap that is filled with this work.

The first step was to derive exact matrix-valued e.o.m.\ for time-dependent correlation functions employing the Zwanzig-Mori projection operator formalism~\cite{Hansen:Theory_of_Simple_Liquids,Goetze:Complex_Dynamics}. The resulting memory kernel depends on all microscopic interactions of the particles among themselves as well as with the boundaries and is generally unknown. Due to the confinement it naturally splits into two relaxation channels. In the collective case, mode-coupling theory~\cite{Goetze:Complex_Dynamics} expresses this kernel as a bilinear functional of the intermediate scattering function, whereas for tagged-particle motion it is a linear functional both of the collective and the incoherent intermediate scattering function. Since it is favourable for the application of mode-coupling approximations, we have extended the idea of irreducible memory kernels~\cite{Cichocki:PhysicaA:1987, Kawasaki:PhysicaA:1995} to multi-channel relaxation for confined liquids. Then also the irreducible memory kernel splits into a component parallel and perpendicular to the confining slit geometry. Following the mode-coupling idea one of our major achievements has been to show that the force kernel and the corresponding coupling coefficients (vertices) coincide  with the Newtonian case. Therefore, it is favourable to use the irreducible memory function instead of the second order memory function~\cite{Pitts:JCP:2000}.

In the framework of Newtonian dynamics the fluctuating force is the time derivative of the momentum-flux tensor (also called the stress tensor), which splits into a kinetic and a potential term. In the MCT functional, however, only the interactions appear as manifested by 
 the direct correlation function and the static structure factor. 

The colloidal case can be considered as a subsystem (solute) suspended in  a solvent also obeying Newtonian dynamics. Then the general equations of motion for the subsystem can be separated into two distinctive parts~\cite{Goetze:Complex_Dynamics}. The first one describes the convective contribution and the interactions due to the potential of the particles within the subsystem. The interactions emanating from interactions between particles with the solvent are included in the second contribution. Consequently, also the fluctuating force splits into two parts, where the first one is translationally invariant with respect to the subsystem and cannot change its total momentum. However, due to the second part momentum of the subsystem is not conserved and can be transferred to the solvent. Nevertheless for cages of neighbouring particles Newton's third law is still fulfilled within the cage, even if momentum conservation is not valid in the colloidal case.

Thus, changing the microscopic dynamics from Newtonian to Brownian motion affects only the short-time behaviour, whereas the structural relaxation described by the MCT functional remains unchanged~\cite{Szamel:PRA:1991,Franosch:PRE_55:1997}. Since the e.o.m.\ have already been conjectured correctly all the resulting conclusions remain valid~\cite{Lang:JStatMech:2013}. For example, the purely relaxating solutions are unique and can be obtained by a convergent iteration scheme. The solutions then correspond to matrix-valued correlation functions, i.e. they are positive-definite and  display non-negative power spectra. Furthermore, the maximum solution of a self-consistent equation for the nonergodicity parameters coincides with the long-time limit of the intermediate scattering function (covariance principle). The solutions for the nonergodicity parameter can be achieved without solving the dynamical equations explicitly (maximum principle). For the generalization of the single-component properties to the matrix-valued theory it is useful to introduce an effective memory kernel~\cite{Lang:JStatMech:2013}. Besides, the phase diagram and the nonergodicity parameters are the same for Newtonian and Brownian dynamics~\cite{Lang:PRE:2012}.

Since for Brownian dynamics the MCT solutions are completely monotone functions also for matrix-valued correlation functions~\cite{Franosch:JStat:2002}, the underlying relaxational dynamics are directly manifested. For confined liquids the numerical computation of the relaxation rate distribution from the MCT solution can be generalized from the bulk case~\cite{Franosch:JPCB:1999}.

Irreducible memory kernels are also present in modified MCT approaches, e.g.\ within the self-consistent generalized Langevin equation theory~\cite{Yeomans-Reyna:PRE_64:2001,Yeomans-Reyna:PRE_76:2007}.  
Our method of adopting the idea of multi-channel relaxation to irreducible memory kernels can be employed generally, e.g.\ to different types of confinement where the density modes can be expanded into a complete set of modes. Of particular interest is the case of quasi-confined liquids where the confinement is realized by periodic boundary conditions in the transverse direction. Then the dynamics are restricted to  a three-dimensional torus implicating that all the matrix-valued quantities become diagonal in the mode index, which makes it more feasible to solve the e.o.m.\ numerically. Static properties of such systems have been investigated recently~\cite{Petersen:JStatMech:2019}, whereas the dynamical solutions will be part of a future publication. 

Multi-channel relaxation is not restricted to confined geometries. For example, the concept can be extended to non-spherical particles like rigid linear~\cite{Scheidsteger:PRE:1997,Franosch:PRE_56:1997,Kaemmerer:PRE:1998,Kaemmerer:PRE:1998b,Scheidsteger:PhilMagB:1998} or arbitrarily shaped molecules~\cite{Fabbian:PRE:2000}. Since binary mixtures display fascinating mixing effects already in the bulk~\cite{Foffi:PRL_91:2003, Goetze:PRE_67:2003}, it would be interesting to extend the MCT to confined mixtures. As already mentioned, multiple relaxation channels have been
 used successfully to describe active microrheology~\cite{Gruber:PRE:2016,Gruber:PHD:2019}.
Another theory incorporating split kernels is adopted for active Brownian particles, where the memory kernel is split into translational and rotational parts~\cite{Liluashvili:PRE:2017}. 
Nevertheless, since in this context orientational degrees of freedom never slow down, the split memory kernel reduces to the translational component quite contrary to our case where all matrix elements are non-vanishing.

\section{Acknowledgements}
We thank Markus Gruber and Matthias Fuchs for useful discussions. This work was supported by the Austrian Science Fund (FWF) under Grant I 2887.

\appendix
\section{Evaluation of the diffusion coefficient matrix element}\label{appendix_diffusion}
The diffusion matrix element of the adjoint Smoluchowski operator Eq.~\eqref{eq:Smoluchowski_adjoint} can be calculated using the definition of the Kubo scalar product Eq.~\eqref{eq:Kubo_scalar} together with the equilibrium distribution Eq.~\eqref{eq:equilibrium_distribution}.
\begin{align}
  \langle\rho_\mu(\vec{q})|\Omega^\dagger\rho_\nu(\vec{q})\rangle =& D_0 \int\mathrm{d}\Gamma \delta\rho_\mu(\vec{q})^*\Big[\sum_{n=1}^N\left(\vec{\nabla}_n-(1/k_B T)\vec{\nabla}_n U\right) \cdot\vec{\nabla}_n\delta\rho_\nu(\vec{q})\Big]
  \psi_{\text{eq}}(\Gamma) \nonumber \\
  =& D_0 \int\mathrm{d}\Gamma \delta\rho_\mu(\vec{q})^*\sum_{n=1}^N\vec{\nabla}_n \cdot \Big[Z^{-1}\exp{(-U/k_B T)}\vec{\nabla}_n \delta\rho_\nu(\vec{q})\Big].
\end{align}
Applying the product rule for the nabla operator we calculate an explicit expression for the diffusion coefficient matrix element
\begin{align}
\langle\rho_\mu(\vec{q})|\Omega^\dagger\rho_\nu(\vec{q})\rangle =& -D_0 \int\mathrm{d}\Gamma  \sum_{n=1}^N \left[\vec{\nabla}_n\delta\rho_\mu(\vec{q})\right]^*\cdot \left[\vec{\nabla}_n\delta\rho_\nu(\vec{q})\right] Z^{-1}\exp{(-U/k_B T)} \nonumber \\
  =& -D_0 \int\mathrm{d}\Gamma  \begin{aligned}[t] &\sum_{\alpha=\parallel,\perp}b^\alpha(q,Q_\mu)b^\alpha(q,Q_\nu) Z^{-1}\exp{(-U/k_B T)}  \\ 
\times &\sum_{n=1}^N 
\left[\exp{(\imag Q_\mu z_n)}e^{\imag\vec{q}\cdot\vec{r}_n}\right]^* \left[\exp{(\imag Q_\nu z_n)}e^{\imag\vec{q}\cdot\vec{r}_n}\right] 
\end{aligned}
\nonumber \\
  =& -D_0 \left(q^2 + Q_\mu Q_\nu\right) \sum_{n=1}^N\langle\exp\left[\imag\left(Q_\nu-Q_\mu\right)z_n\right]\rangle \nonumber \\
  =& -D_0 \left(q^2+Q_\mu Q_\nu\right)A n_{\mu-\nu}^*,
\end{align}
where obviously $\vec{\nabla}_n\delta\rho_{\mu}(\vec{q})=\vec{\nabla}_n\rho_{\mu}(\vec{q})$ is valid.
In the last line $\sum_{n=1}^N\langle\exp\left[\imag\left(Q_\nu-Q_\mu\right)z_n\right]\rangle=\langle\rho_{\nu-\mu}(0)\rangle=An_{\nu-\mu}=An_{\mu-\nu}^*$ has been used.

\section{Evaluation of the fluctuating force}\label{appendix_fluctuatingforce}
We also calculate the 'fluctuating force' explicitly
\begin{align}
 \mathcal{Q}_{\rho}\Omega^\dagger \ket{\rho_\mu(\vec{q})} =& \mathcal{Q}_{\rho} D_0 \sum_{n=1}^N \left(\vec{\nabla}_n-\frac{1}{k_BT}\vec{\nabla}_n U\right)\cdot \vec{\nabla}_n
  e^{\imag \vec{q}\cdot\vec{r}_n} \exp{(\imag Q_{\mu} z_n)} \nonumber \\
  =&  -\mathcal{Q}_{\rho} D_0 \sum_{\alpha=\parallel,\perp}  b^\alpha(q,Q_{\mu})^2 \sum_{n=1}^N e^{\imag \vec{q}\cdot\vec{r}_n}\exp{(\imag Q_\mu z_n)}\nonumber\\
  &- \imag \frac{D_0}{k_B T}\sum_{\alpha=\parallel,\perp}b^\alpha(q,Q_\mu)\mathcal{Q}_{\rho}\sum_{n=1}^N  b^\alpha\left(\frac{\partial U}{\partial \vec{r}_n},\frac{\partial U}{\partial z_n}\right)e^{\imag\vec{q}\cdot\vec{r}_n} \exp{(\imag Q_{\mu}z_n)},
\end{align}
with $\mathcal{Q}_\rho=1-\mathcal{P}_\rho$. Inserting the corresponding projector from Eq.~\eqref{eq:projector}
\begin{align}
 \mathcal{Q}_{\rho}\Omega^\dagger \ket{\rho_\mu(\vec{q})} =& -(q^2+Q_{\mu}^2)D_0\Bigg(\delta\rho_{\mu}(\vec{q})+\braket{\rho_{\mu}(\vec{q})} \nonumber \\
  &-\frac{1}{N}\sum_{\vec{q}^{\,\prime}}\sum_{\mu^\prime\nu^\prime}\delta\rho_{\mu^\prime}(\vec{q}^{\,\prime})\left[\mathbf{S}^{-1}(q^\prime)\right]_{\mu^\prime\nu^\prime}\braket{\delta\rho_{\nu^\prime}(\vec{q}^{\,\prime})^*
  \delta\rho_{\mu}(\vec{q})}\Bigg) \nonumber \\
  &-\imag \frac{D_0}{k_B T}\sum_{\alpha=\parallel,\perp}b^\alpha(q,Q_\mu)\mathcal{Q}_{\rho}\sum_{n=1}^N b^\alpha\left(\frac{\partial U}{\partial \vec{r}_n},\frac{\partial U}{\partial z_n}\right)e^{\imag\vec{q}\cdot\vec{r}_n}\exp{(\imag Q_{\mu}z_n)},
\end{align}
and introducing
\begin{align}
 \delta F_{\mu}^\alpha(\vec{q}) = -\sum_{n=1}^Nb^\alpha\left(\frac{\partial U}{\partial\vec{r}_n},\frac{\partial U}{\partial z_n}\right) e^{\imag\vec{q}\cdot\vec{r}_n}
  \exp{(\imag Q_{\mu}z_n)} + \imag k_B T b^\alpha(q,Q_\mu)\braket{\rho_{\mu}(\vec{q})},
\end{align}
with the average value of the fluctuating force (c.f.\ Eq.~\eqref{eq:average_dynamical_variable})
\begin{align}
 \braket{F_{\mu}^{\alpha}(\vec{q})} &= -k_B T\Big\langle\sum_{n=1}^N\vec{\nabla}_n\left(e^{\imag\vec{q}\cdot\vec{r}_n}
  \exp(\imag Q_{\mu}z_n)\right)\big\rangle = -\imag k_B T b^\alpha(q,Q_\mu)\braket{\rho_{\mu}(\vec{q})},
\end{align}
the expression simplifies to
\begin{align}
\mathcal{Q}_{\rho}&\Omega^\dagger \ket{\rho_\mu(\vec{q})}= \imag\frac{D_0}{k_B T} \sum_{\alpha}b^\alpha(q,Q_{\mu}) \mathcal{Q}_{\rho}\ket{F^{\alpha}_{\mu}(\vec{q})}.
\end{align}

\section{Evaluation of the overlap matrix element}\label{appendix_overlap}
Here we evaluate the scalar product $\braket{F_{\mu}^{\alpha}(\vec{q})|\mathcal{Q}_{\rho}\rho_{\mu_1}(\vec{q}_1)\rho_{\mu_2}(\vec{q}_2)}$. Using $\mathcal{Q}_{\rho}=1-\mathcal{P}_{\rho}$ there are four contributions which are considered separately:
\begin{align}\label{eq:overlap}
 \langle\delta F_{\mu}^{\alpha}(\vec{q})^*\mathcal{Q}_{\rho}\delta\rho_{\mu_1}(\vec{q}_1)\delta\rho_{\mu_2}(\vec{q}_2)\rangle 
  =& \braket{F_{\mu}^{\alpha}(\vec{q})^*\delta\rho_{\mu_1}(\vec{q}_1)\delta\rho_{\mu_2}(\vec{q}_2)}-\braket{F_{\mu}^{\alpha}(\vec{q})^*\mathcal{P}_{\rho}\delta\rho_{\mu_1}(\vec{q}_1)\delta\rho_{\mu_2}(\vec{q}_2)} \nonumber\\
  &-\imag k_B T b^\alpha(q,Q_\mu)\braket{\rho_{\mu}(\vec{q})^*}\braket{\delta\rho_{\mu_1}(\vec{q}_1)\delta\rho_{\mu_2}(\vec{q}_2)}
  \nonumber\\
  &+\imag k_B T b^\alpha(q,Q_\mu)\braket{\rho_{\mu}(\vec{q})^*}
  \braket{\mathcal{P}_{\rho}\delta\rho_{\mu_1}(\vec{q}_1)\delta\rho_{\mu_2}(\vec{q}_2)}.
\end{align}
For the first term
\begin{align}\label{eq:overlap_term1}
 \langle F_{\mu}^{\alpha}(\vec{q})^*\delta\rho_{\mu_1}(\vec{q}_1)\delta\rho_{\mu_2}(\vec{q}_2)\rangle 
 = -\Big\langle\sum_{n=1}^Nb^\alpha\left(\frac{\partial U}{\partial\vec{r}_n},\frac{\partial U}{\partial z_n}\right) e^{-\imag\vec{q}\cdot\vec{r}_n}\exp{(-\imag Q_{\mu}z_n)} \delta\rho_{\mu_1}(\vec{q}_1)\delta\rho_{\mu_2}(\vec{q}_2)\Big\rangle,
\end{align}
we use the fact that for an arbitrary dynamical variable $A$ its average weighted by an internal force is given by
\begin{align}\label{eq:average_dynamical_variable}
 \braket{(\vec{\nabla}_n U)A} =& \int\mathrm{d}\Gamma \psi_{\text{eq}}\left[\vec{\nabla}_n U(\Gamma)\right]A(\Gamma)
  \nonumber \\
  =& \int\mathrm{d}\Gamma \frac{1}{Z}e^{-U(\Gamma)/k_B T}\left[\vec{\nabla}_n U(\Gamma)\right]A(\Gamma) \nonumber \\
  =& -\int\mathrm{d}\Gamma \vec{\nabla}_n \cdot \left(\frac{k_B T}{Z}e^{-U(\Gamma)/k_B T}A(\Gamma)\right) +\int\mathrm{d}\Gamma \frac{k_B T}{Z} e^{-U(\Gamma)/k_B T} \vec{\nabla}_n A(\Gamma) \nonumber \\
  =& \int\mathrm{d}\Gamma \frac{k_B T}{Z} e^{-U(\Gamma)/k_B T}\vec{\nabla}_n A(\Gamma) \nonumber \\
  =& k_B T\braket{\vec{\nabla}_n A}.
\end{align}
Plugging this result into Eq.~\eqref{eq:overlap_term1} the first contribution is given by
\begin{align}
\langle F_{\mu}^{\alpha}(\vec{q})^*&\delta\rho_{\mu_1}(\vec{q}_1)\delta\rho_{\mu_2}(\vec{q}_2)\rangle \nonumber \\
  =& -k_B T \Big\langle\sum_{n=1}^N b^\alpha\left(\frac{\partial}{\partial\vec{r}_n},\frac{\partial}{\partial z_n}\right)  
  \left( e^{-\imag\vec{q}\cdot\vec{r}_n}\exp{(-\imag Q_{\mu}z_n)}
  \delta\rho_{\mu_1}(\vec{q}_1)\delta\rho_{\mu_2}(\vec{q}_2)\right)\Big\rangle \nonumber\\
  =& -k_B T \Big\langle\sum_{n=1}^N  \delta\rho_{\mu_1}(\vec{q}_1)\delta\rho_{\mu_2}(\vec{q}_2)b^\alpha\left(\frac{\partial}{\partial\vec{r}_n},\frac{\partial}{\partial z_n}\right) \left( e^{-\imag\vec{q}\cdot\vec{r}_n}\exp{(-\imag Q_{\mu}z_n)}\right) 
 \Big\rangle \nonumber \\
  &- k_B T \Big[\Big\langle\sum_{n=1}^N e^{-\imag\vec{q}\cdot\vec{r}_n}\exp{(-\imag Q_{\mu}z_n)}\delta\rho_{\mu_2}(\vec{q}_2) b^\alpha\left(\frac{\partial}{\partial\vec{r}_n},\frac{\partial}{\partial z_n}\right) \left(
  \delta\rho_{\mu_1}(\vec{q}_1)\right)\Big\rangle + (1 \leftrightarrow 2)\Big]\nonumber\\
  = &\imag k_B T b^{\alpha}(q,Q_\mu)\Big[ \braket{\rho_{\mu}(\vec{q})|\rho_{\mu_1}(\vec{q}_1)\rho_{\mu_2}(\vec{q}_2)} +  \braket{\rho_{\mu}(\vec{q})^*}\braket{\delta\rho_{\mu_1}(\vec{q}_1)\delta\rho_{\mu_2}(\vec{q}_2)}\Big] \nonumber\\
  &-\imag k_B T \Big[ b^{\alpha}(\hat{\vec{q}}\cdot\vec{q}_1,Q_{\mu_1}) 
  \braket{\rho_{\mu-\mu_1}(\vec{q}-\vec{q}_1)|
  \rho_{\mu_2}(\vec{q}_2)} + (1 \leftrightarrow 2) \Big],
\end{align}
where $\braket{\rho_{\mu-\mu_1}(\vec{q}-\vec{q}_1)^*\delta\rho_{\mu_2}(\vec{q}_2)}=\braket{\delta\rho_{\mu-\mu_1}(\vec{q}-\vec{q}_1)^*\delta\rho_{\mu_2}(\vec{q}_2)}$ has been used, since the second contribution $\braket{\rho_{\mu-\mu_1}(\vec{q}-\vec{q}_1)}\braket{\delta\rho_{\mu_2}(\vec{q}_2)}$ vanishes. Abbreviating the static three-point correlation function by
\begin{equation}
 S_{\sigma,\mu_1\mu_2}(\vec{q},\vec{q}_1\vec{q}_2) = \frac{1}{N}\braket{
  \delta\rho_{\sigma}(\vec{q})^*\delta\rho_{\mu_1}(\vec{q}_1)\delta\rho_{\mu_2}(\vec{q}_2)},
\end{equation}
we arrive at the final expression for the first term in Eq.~\eqref{eq:overlap}
\begin{align}
\langle F_{\mu}^{\alpha}(\vec{q})^*&\delta\rho_{\mu_1}(\vec{q}_1)\delta\rho_{\mu_2}(\vec{q}_2)\rangle \nonumber \\
  = &\imag k_B T\Big[ b^{\alpha}(q,Q_\mu) N S_{\mu,\mu_1\mu_2}(\vec{q},\vec{q}_1\vec{q}_2)\delta_{\vec{q},\vec{q}_1+\vec{q}_2} + \braket{\rho_{\mu}(\vec{q})^*}
  \braket{\delta\rho_{\mu_1}(\vec{q}_1)\delta\rho_{\mu_2}(\vec{q}_2)}\Big] \nonumber\\
  &- \imag k_B T \Big[ b^{\alpha}(\hat{\vec{q}}\cdot\vec{q}_1,Q_{\mu_1}) N S_{\mu-\mu_1,\mu_2}(q_2)
  \delta_{\vec{q},\vec{q}_1+\vec{q}_2} +(1 \leftrightarrow 2) \Big],
\end{align}
where the second term cancels with the third term in Eq.~\eqref{eq:overlap}.

For the next contribution we also have to evaluate the projection on the density modes. Therefore,
\begin{align}
 \langle F_{\mu}^{\alpha}(\vec{q})^*&\mathcal{P}_{\rho}\delta\rho_{\mu_1}(\vec{q}_1)\delta\rho_{\mu_2}(\vec{q}_2)\rangle \nonumber \\
  =& -\Big\langle\sum_{n=1}^N b^\alpha\left(\frac{\partial U}{\partial\vec{r}_n},\frac{\partial U}{\partial z_n}\right) e^{-\imag\vec{q}\cdot\vec{r}_n}\exp{(-\imag Q_{\mu}z_n)} \mathcal{P}_{\rho}\delta\rho_{\mu_1}(\vec{q}_1)\delta\rho_{\mu_2}(\vec{q}_2)\Big\rangle \nonumber\\
  =& -\Big\langle\sum_{n=1}^N b^\alpha\left(\frac{\partial U}{\partial\vec{r}_n},\frac{\partial U}{\partial z_n}\right)e^{-\imag\vec{q}\cdot\vec{r}_n}\exp{(-\imag Q_{\mu}z_n)} 
  \frac{1}{N}
  \sum_{\vec{q}^{\,\prime}}\sum_{\kappa\sigma}\delta\rho_{\kappa}(\vec{q}^{\,\prime})\Big\rangle \nonumber \\ &\times \left[\mathbf{S}^{-1}(q^\prime)\right]_{\kappa\sigma}\braket{\rho_{\sigma}(\vec{q}^{\,\prime})|\rho_{\mu_1}(\vec{q}_1)\rho_{\mu_2}(\vec{q}_2)} \nonumber\\
  =& -\frac{k_B T}{N}\sum_{\kappa\sigma}\sum_{\vec{q}^{\,\prime}}\Big\langle\sum_{n=1}^N b^\alpha\left(\frac{\partial }{\partial\vec{r}_n},\frac{\partial }{\partial z_n}\right)
  \left( e^{-\imag\vec{q}\cdot\vec{r}_n}\exp{(-\imag Q_{\mu}z_n)}\delta\rho_{\kappa}(\vec{q}^{\,\prime})\right)\Big\rangle \nonumber\\
  &\times \left[\mathbf{S}^{-1}(q^\prime)\right]_{\kappa\sigma}\braket{\rho_{\sigma}(\vec{q}^{\,\prime})|\rho_{\mu_1}(\vec{q}_1)
  \rho_{\mu_2}(\vec{q}_2)} \nonumber\\
  =& \imag\frac{k_B T}{N}\sum_{\kappa\sigma}\sum_{\vec{q}^{\,\prime}}\Big(b^{\alpha}(q,Q_\mu)
  \braket{\rho_{\mu}(\vec{q})^*\delta\rho_{\kappa}(\vec{q}^{\,\prime})} 
  -b^{\alpha}(\hat{\vec{q}}\cdot\vec{q}^{\,\prime},
  Q_{\kappa})\braket{\rho_{\kappa-\mu}(\vec{q}^{\,\prime}-\vec{q})}\Big) \nonumber \\
  &\times\left[\mathbf{S}^{-1}(q^\prime)\right]_{\kappa\sigma}\braket{\rho_{\sigma}(\vec{q}^{\,\prime})|\rho_{\mu_1}(\vec{q}_1)\rho_{\mu_2}(\vec{q}_2)} \nonumber\\
  =& \imag k_B T N \delta_{\vec{q},\vec{q}_1+\vec{q}_2}\Big(b^{\alpha}(q,Q_\mu) S_{\mu,\mu_1\mu_2}(\vec{q},\vec{q}_1\vec{q}_2) \nonumber\\
  &-\sum_{\kappa\sigma}b^{\alpha}(q,Q_{\kappa})\frac{n_{\mu-\kappa}^*}{n_0}\left[\mathbf{S}^{-1}(q)\right]_{\kappa\sigma}S_{\sigma,\mu_1\mu_2}(\vec{q},\vec{q}_1\vec{q}_2)\Big).
\end{align}
In the last line we used again the definition of the triple correlation function and $n_0=N/A$. Finally, it is easy to show that the last term in Eq.~\eqref{eq:overlap} vanishes
\begin{align}
 \langle\mathcal{P}_{\rho}&\delta\rho_{\mu_1}(\vec{q}_1)\delta\rho_{\mu_2}(\vec{q}_2)\rangle 
  =\frac{1}{N}\sum_{\vec{q}}\sum_{\mu\nu}
  \braket{\delta\rho_{\mu}(\vec{q})}\left[\mathbf{S}^{-1}(q)\right]_{\mu\nu} 
  \braket{\delta\rho_{\nu}(\vec{q})\delta\rho_{\mu_1}(\vec{q}_1)\delta\rho_{\mu_2}(\vec{q}_2)}=0.
\end{align}
Combining everything one finds the explicit representation of the overlap matrix element
\begin{align}
 \langle F_{\mu}^{\alpha}(\vec{q})|\mathcal{Q}_{\rho}\rho_{\mu_1}(\vec{q}_1)\rho_{\mu_2}(\vec{q}_2)\rangle 
= -&\imag N k_B T \delta_{\vec{q},\vec{q}_1+\vec{q}_2}\Big[b^{\alpha}(\hat{\vec{q}}\cdot\vec{q}_1,Q_{\mu_1})
  S_{\mu-\mu_1,\mu_2}(q_2) + (1 \leftrightarrow 2) \nonumber \\
 &- \sum_{\kappa\sigma}\frac{n_{\mu-\kappa}^*}{n_0} b^{\alpha}(q,Q_\kappa)\left[\mathbf{S}^{-1}(q)\right]_{\kappa\sigma} 
   S_{\sigma,\mu_1\mu_2}(\vec{q},\vec{q}_1\vec{q}_2)\Big].
\end{align}

\bibliographystyle{tfq}

\end{document}